\documentclass{PoS}
\usepackage{units}
\usepackage[numbers]{natbib}

\newcommand*\aap{A\&A}

\newcommand*\ao{Appl.~Opt.}

\newcommand*\apjl{ApJ}

\newcommand*\apjs{ApJS}

\newcommand{\thispsr}{PSR~B1706$-$44}
\newcommand{\hess}{H.E.S.S.}
\newcommand{\hessII}{\hess{}~II}
\newcommand{\fermi}{{\emph{Fermi}}}
\newcommand{\gr}{$\gamma$-ray}
\newcommand{\grs}{\gr{}s}

\title{Detection of sub-100 GeV \gr{} pulsations from \thispsr{} with \hess{}}

\ShortTitle{Detection of \thispsr{} with \hess{}}

\author{\speaker{M. Spir-Jacob}$^1$, A.~Djannati-Ata\"i$^1$, L.~Mohrmann$^2$, G.~Giavitto$^3$, B.~Khelifi$^1$, B.~Rudak$^4$, C.~Venter$^5$, R.~Zanin$^6$ for the H.E.S.S. Collaboration$^7$ \\

$^1$ APC, Universit\'{e} de Paris, CNRS/IN2P3, CEA/Irfu, Observatoire de Paris, 75205 Paris, France\\

$^2$ Friedrich-Alexander-Universit\"at Erlangen-N\"urnberg, Erlangen Centre for Astroparticle Physics, Erwin-Rommel-Str. 1, D 91058 Erlangen, Germany \\

$^3$ DESY, D-15738 Zeuthen, Germany \\

$^4$ Nicolaus Copernicus Astronomical Center, Polish Academy of Sciences, ul. Bartycka 18, 00-716 Warsaw, Poland \\

$^5$ Centre for Space Research, North-West University, Potchefstroom 2520, South Africa \\

$^6$ Max-Planck-Institut f\"ur Kernphysik, P.O. Box 103980, D 69029 Heidelberg, Germany \\

$^7$ for collaboration list see PoS(ICRC2019)1177\\
}

\abstract{We report on the detection of pulsations from \thispsr{} based on 28.3 hours of observations with the H.E.S.S. II array with CT5 
in monoscopic mode. The lightcurve is similar to that obtained with the \fermi-LAT above 15~GeV and
the pulsations exhibit a steep spectrum with index $\sim -3.8$ in the sub 20~GeV to sub-100~GeV energy range.
While a significant signal of $\sim 1000$ events is detected at energies $\sim 70$~GeV, it is not possible to either confirm or 
rule out a power-law behaviour of \thispsr{} spectrum in this range.}

\FullConference{36th International Cosmic Ray Conference -ICRC2019-\\
  July 24th - August 1st, 2019\\
  Madison, WI, U.S.A.}

\begin{document}

\section{Introduction}

The 28 m equivalent diameter telescope (CT5), added in 2012 to the core of the H.E.S.S. array of four 12 m 
diameter imaging atmospheric Cherenkov telescopes (CT1-4) has allowed to reach sub-20 GeV energies in monoscopic 
mode for observations of pulsars \cite{2017AIPC.1792d0028D,VelaMonoPaper},
 thus bridging the gap 
to satellite-based \gr{} instruments. 

\thispsr{} is the third brightest \gr{} emitting pulsar after the Vela and Geminga pulsars\cite{Abdo20132PC}, which have been detected from the ground along with the Crab pulsar \cite{VelaMonoPaper,veritasvhecrab, Aleksic2011,Ansoldi2016, 2019ICRCmagicgeminga}.
Its spin-down power of $\dot E= 3.4 \times 10^{36}$~erg/s 
and age, $\sim 1.7 \times 10^4$ years are very similar to those of the Vela pulsar. 
Despite its rather large distance of $\sim$2.3 kpc as compared to Vela (294 pc), 
the \gr{} luminosity of \thispsr{} at 10 GeV is only 3 times lower than Vela, and as such,
it constituted a promising target for detection from ground.


\grs{} were first detected from  \thispsr{} with the COS-B satellite in 1981
\cite{swanenburg1981}, classified as an unidentified source. It was only identified a decade later as a 102 ms
pulsar with the Parkes radio telescope \cite{johnston1992}. 
\thispsr{} has since been detected with EGRET \cite{thompson92}, \textit{Chandra} \cite{gotthelf2002},
AGILE \cite{pittori2009} and \fermi-LAT \cite{Abdo2010EGRET}.
The synchrotron nebula around \thispsr{} \cite{frail1994} displays a surprisingly low radio flux 
in comparison with other radio pulsar wind nebulae (PWNe) \cite{giacani2001}, and in X-rays, \textit{ROSAT} 
and \textit{ASCA} have revealed a structure of a torus and a jet \cite{finley1998}.
In VHE \grs{}, an extended source of Gaussian width $0.29^\circ$ was discovered by \hess{} above 
600~GeV \cite{HESSb1706}, of which the association with \thispsr{} is likely but not firmly established.  
     
\section{Observations and data analysis} 
\subsection{\hess}

H.E.S.S. observations were carried out during
two campaigns 
in 2013 and 2015.
They were made in
wobble mode with an offset range of  $0.2^\circ-0.7^\circ$, and an average zenith angle of  $24.5^\circ$. 
After quality selection for smooth telescope operation and good weather conditions, 
28.3 hours of data out of the 38 hours of observations were kept. 

The monoscopic analysis pipeline used for this detection is the same as the one originally developed and validated on Vela pulsar data with CT5 \cite{VelaMonoPaper}. It was used to reconstruct the shower direction, impact and energy of the primary \grs{}, based on the recorded shower images. The images were obtained 
after calibration and image cleaning. 
To compute the phase of each event, the time stamps provided by the central trigger system of H.E.S.S. 
were folded using the \texttt{Tempo2} package \cite{TempoGeneral} with an ephemeris using both LAT timing \cite{Kerr:2015tva} and radio data from the Parkes Radio Telescope \cite{Weltevrede2010}, valid between 
$22^{\rm{nd}}$ of July 2007 till $11^{\rm{th}}$ of September 2015\footnote{Kindly provided by Matthew Kerr and David A. Smith.}.

To discriminate between photons and hadrons, a boosted decision tree (BDT), trained 
on \gr{} Monte Carlo (MC) simulations for the signal and on 
real data for the background, was used. Background was additionally rejected through a 
spatial cut at a 68\%-containment radius ($0.3^\circ$) 
and by a selection in phase. For the latter, 
ON- and OFF-phase ranges were defined based on the \textit{Fermi}-LAT phasogram (see below)
and a maximum likelihood-ratio test \cite{lima1983} was applied to compute the significance of the signal. 

The energy spectrum for \hess{} data was derived using a  maximum
likelihood fit within a forward-folding scheme~\cite{Piron2001}.
The instrument response functions (IRFs) were computed through extensive MC simulations as a
function of the energy, zenith, and azimuthal angles of the telescope
pointing direction, the impact parameter of showers, and the
configuration of the telescope for each observing period.  

\begin{figure}[th]
\centering
\includegraphics[width=\columnwidth]{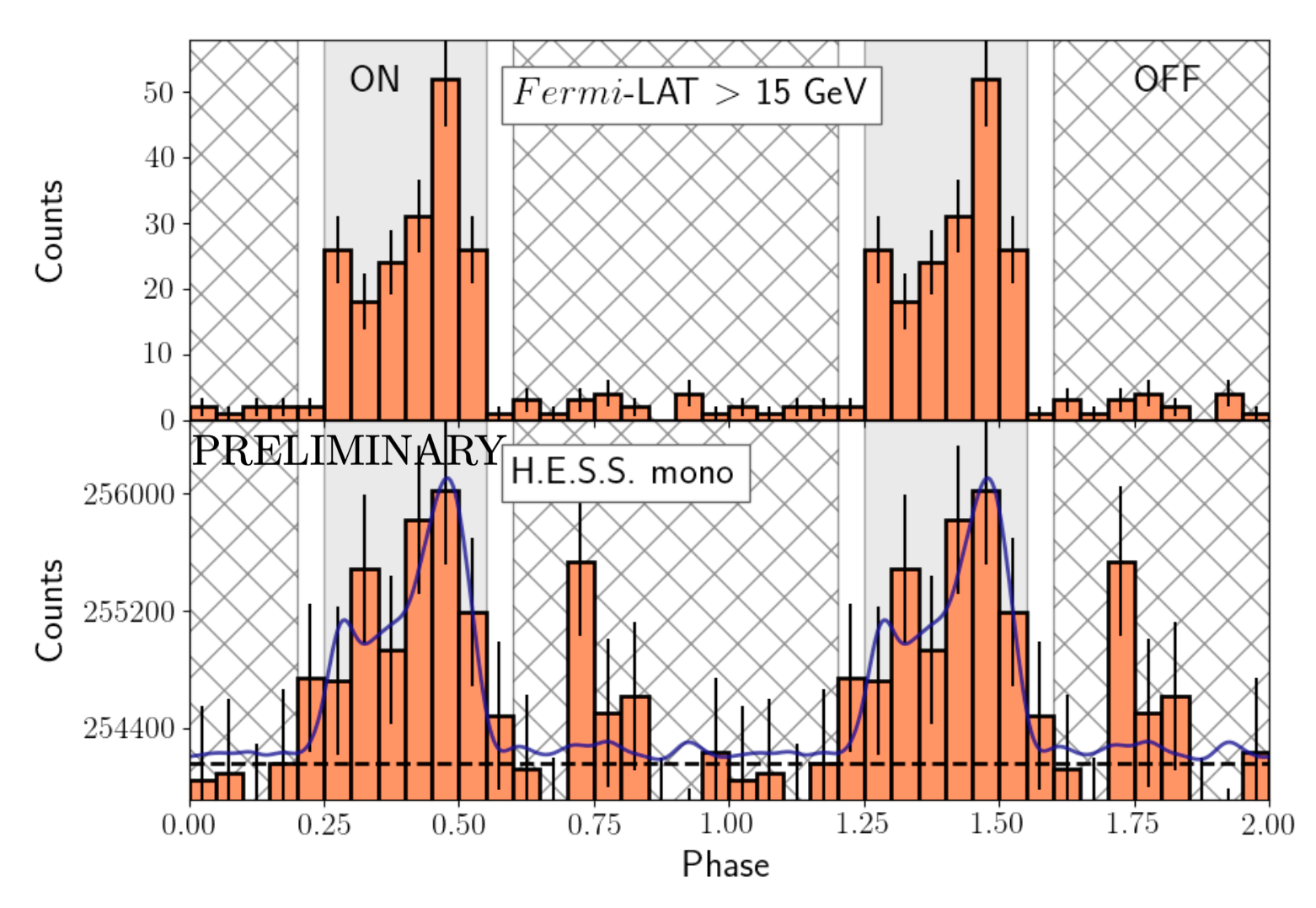}
\caption{Histogram of phases for two periods with \textit{Fermi}-LAT data $>$ 15 GeV (top) and H.E.S.S. II with CT5 in monoscopic mode data (bottom). 
The hashed box represents the OFF-pulse region and the grey box the ON-pulse region. The empty (or white) area between both isn't considered as part of the ON nor OFF region. The dashed line shows the average level of background evaluated in the OFF-pulse region. From the $Fermi$-LAT data above 15 GeV, we derive a KDE (variance of 0.025), 
represented by the blue curve and used for a maximum-likelihood ratio test on the \hessII-CT5 data.}
\label{phaso}
\end{figure}

\subsection{\fermi-LAT}

62 months of \textit{Fermi} data from 
2008 to 
2013 were used to derive the phasogram and 
phased-resolved spectra above two energy thresholds of 100~MeV and 10~GeV.

Events were selected 
from the P8 Source class (event class = 128, event type =3) within a 
region of interest (ROI) of $\unit[10]{^{\circ}}$ radius around the position of the  pulsar, and
\texttt{P8R2\_SOURCE\_V6} IRFs were used.
Only $\gamma$-ray events 
with reconstructed zenith angles smaller than $\unit[90]{^{\circ}}$ were selected 
in order to reduce contamination by $\gamma$-rays from Earth's limb.


The pulsar phase was computed for selected events 
using the \texttt{Tempo2} \fermi{} plug-in \cite{Ray_2011} and the same ephemeris as 
that used for the \hess{} data. 

An additional selection cut $\theta_{\rm max}=0.6^\circ$ was applied 
on the angular distance of each photon to the pulsar 
position for generation of the light curves. This cut value is slightly smaller than the 
68\% and 95\% containment radii of the \fermi-LAT at 1 and 10~GeV, respectively, and allows us
to retain a large number of highest energy photons, while limiting the background in the $1-10$~GeV range.  

The spectral analysis was done using \texttt{gtlike} tool, 
the Galactic diffuse emission model, {\footnotesize \texttt{gll\_iem\_v06.fits}}, and
isotropic diffuse model, {\footnotesize \texttt{iso\_P8R2\_SOURCE\_V6\_v06.txt}}.
All sources from the \fermi-LAT third source catalogue (3FGL) \cite{2015ApJS..218...23A}, within a region of
$\unit[20]{^{\circ}}$ radius centred on the pulsar position were added to
the source model, while parameters for sources outside the ROI were
fixed during the fit. 

\section{Results} 
\label{sec::detection}
\subsection{Light Curves}
\label{lightcurve}

Figure \ref{phaso} shows the phasogram of \thispsr{} for the
\textit{Fermi}-LAT data above 15 GeV (on top) and for the \hess{} data (on
the bottom). Two periods are displayed for better readability.  
The light curve of \thispsr{} at lower energies, e.g. 1 GeV,
consists of two peaks spanning in the [0.25-0.55] phase range and connected with a 
large-amplitude bridge \cite{Abdo2010EGRET}. 
The [0.25-0.55] and [0.6-0.2] intervals were subsequently defined 
as the ON- and OFF-phase ranges, respectively. 

The H.E.S.S. II-CT5 light curve contains 5 091 420 events, with 1 532 177 in the ON-phase and 3 050 011 in 
the OFF-phase, which corresponds to a 7171.5$\pm$ 1515 excess. The source is detected 
at a significance level of 4.74$\sigma$ (Li\&Ma test \cite{lima1983}).
An alternative pipeline (\cite{murach2015}) was used as a cross-check
and validated the detection, at a slightly lower significance.
We note that the phase bins near phase 0.75 exhibit some excess but at a low significance level when taking into account 
the trials. 
The probability density function (PDF) of the \fermi-LAT light curve is derived through a KDE (Kernel Density Estimator) with variance 0.025, and used
in a maximum-likelihood ratio test on the \hess{} data. The test significance reaches a value of 4.6$\sigma$ for 8139 signal events, which is compatible with the Li \& Ma test results. The fact that the number of signal events found with the maximum likelihood ratio test is larger than the number of excess events based on the ON and OFF-pulse zones can be due to the contribution of some events near phase 0.75.

\subsection{Spectra}
\label{spectrasec}

\begin{figure}[h]
\centering
\includegraphics[width=\columnwidth]{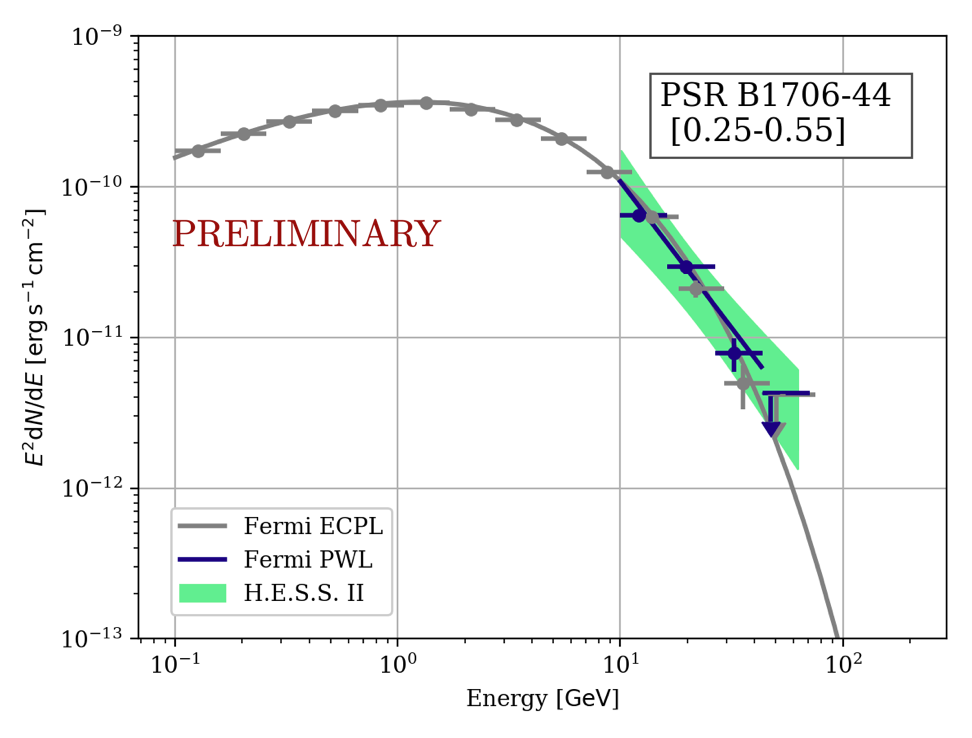}
\caption{PSR B1706-44 spectral energy distribution. The grey and blue flux points are obtained from 5 years of \textit{Fermi} data. 
Two fits are plotted: the power law with a sub-exponential cutoff above 100 MeV (grey) and another power law for the tail of the emission 
above 10 GeV (blue). The green box is derived from 28.3 hours of H.E.S.S. II-CT5 data and includes systematic errors (see text).}
\label{spectra}
\end{figure}

The spectra obtained with both instruments and with the phase definitions given above are shown in Fig. \ref{spectra}.
For the \textit{Fermi}-LAT data, phase-resolved spectra were derived first above 100~MeV,
assuming a power law with an exponential cut-off (ECPL): \\ 
\indent ${\rm d}N(E)/{\rm d}E = N_0 \left({E}/{E_0}\right)^{-\Gamma} \exp\left[-\left({E}/{E_\mathrm{c}}\right)^b\right].$ \\
The best-fit values 
obtained are $ N_0= (1.05\pm0.05^{\rm stat}) \times 10^{-9}$ {MeV}$^{-1}$cm$^{-2}$s$^{-1}$, $\Gamma=1.19\pm0.01$, $b=0.48\pm0.01$ and $E_c = 403 \pm 10$ MeV at a
reference energy $E_0$ of 1~GeV.
The fit of a power law above 10~GeV yielded in turn an index $\Gamma_{\rm LAT}=3.9\pm 0.1$ and a flux normalisation 
at the reference energy of 20~GeV $ N_0= (4.4\pm0.3^{\rm stat}) \times 10^{-8}$ {MeV}$^{-1}$cm$^{-2}$s$^{-1}$.
The fit of a parabola model (LPB, ${\rm d}N(E)/{\rm d}E  = \Phi_0 \left({E}/{E_0}\right)^{-\Gamma_{\rm LPB}-\beta\ \ln({E}/{E_0})}$)
was performed to investigate curvature above 10~GeV,  
and resulted in $\Gamma_{\rm LPB}= 4.1\pm0.2$ and  $\beta=0.5\pm0.4$. The likelihood ratio between the parabola model and the power law yields a significance of only $1.4\sigma$ in favor of the curvature. This, however, does not preclude the exponential cut-off found with the larger energy range fit, although one might 
be in the same situation as that of the Crab \cite{Ansoldi2016}, i.e. an apparent curvature due to lack of statistics.

For the H.E.S.S. II-CT5 data, a power-law fit 
resulted in an index $\Gamma_{\rm HESS} = 3.76 \pm 0.36^{\rm stat}$, a normalization
$\Phi_0^{\rm HESS}=(4.3\pm0.9^{\rm stat}) \times 10^{-8}$ {TeV}$^{-1}$cm$^{-2}$s$^{-1}$, at the reference energy $E_0=20$~GeV, and with
decorrelation energy $E_{\rm d}=21.5$ GeV. It was shown in~\cite{VelaMonoPaper} that the threshold of CT5 is close to 10 GeV.
Here, again, the CT5 data fit results are in full agreement with the ones obtained from the LAT above 10~GeV.


The fit of a parabola model was not attempted due to lack of statistics and rather low signal-to-noise ratio.   
The highest energy bin in the CT5 data, 54 to 225 GeV, displays an excess of 2782 events at a significance level of 2.5$\sigma$.

Due to the large bias in energy reconstruction (see \cite{VelaMonoPaper}), the average energy in this bin differs from a simple weighted 
mean taking into account the spectral index. An evaluation using 
a simulated spectrum with parameters matching those of the \fermi-LAT power law above 10 GeV predicts that 60\% of events with $<E>$=62.7~GeV and with a dispersion of 37 GeV lie in that bin. The confidence box of the \hess{} SED, shown in Fig~\ref{spectra} is hence limited to 62.7~GeV.
The box includes the systematic errors obtained with the procedure described in~\cite{VelaMonoPaper}, except that the energy scale uncertainty
used here is +5\% instead of +8\%. The latter value was an upper limit for the relative energy scale offset between \textit{Fermi}-LAT and H.E.S.S. II-CT5.
The correction to +5\% is due to the LAT recalibration 
on electrons~\cite{LatElectronRecalibration} 
which brought the LAT scale 3\% down in energy, i.e. closer to CT5 scale.  


\section{Summary}

A significant pulsed signal from \thispsr{} has been detected with 28.3 hours of observations with 
\hessII-CT5 in monoscopic mode. 
This is the fourth detection of a pulsar from the ground, after the Crab, Vela and Geminga pulsars. 

The phasogram obtained with H.E.S.S. II-CT5 is similar to that of \fermi-LAT above 15~GeV, i.e. the peak P2 is clearly dominant over P1, in 
continuity with the trend seen at lower energies \cite{Abdo2010EGRET}.
The comparison of the CT5 spectrum with that obtained from 5 years of \textit{Fermi}-LAT data above 10~GeV shows 
a very good agreement. The pulsed spectrum of \thispsr{} above 10~GeV is shown to be very steep (index$\sim-3.8$ to $3.9$) with both 
\fermi-LAT and H.E.S.S. II-CT5 and similar to the Vela pulsar~\cite{VelaMonoPaper}. However, 
the lack of statistics in either instrument data prevents any conclusion on the absence or presence of a spectral curvature/cut-off.
The question hence remains whether PSR B1706-44's spectrum behaves like the Vela PSR where an indication 
of curvature at a level $> 3 \sigma$ was found independently with both \fermi{} and \hessII-CT5 \cite{VelaMonoPaper}, or as 
the Crab pulsar where a soft and tail-like extension of the \fermi-LAT spectrum was found above 100~GeV and 
extending up to 1~TeV~\cite{veritasvhecrab, Aleksic2011,Ansoldi2016}.
Measuring the behaviour of the tail of the spectrum in the tens of GeV range should bring further insights
into mechanisms at play in young  pulsars, including those in the VHE regime. This would require further
observations with H.E.S.S. or CTA.

\section{Acknowledgements}

See: \texttt{https://www.mpi-hd.mpg.de/hfm/HESS/pages/publications/auxiliary/} \texttt{HESS-Acknowledgements-2019.html}

\bibliographystyle{aa}
\bibliography{main}
\end{document}